\documentstyle[sprocl]{article}

\def\Journal#1#2#3#4{{#1} {\bf #2}, #3 (#4)}

\def\NPB{{\em Nucl. Phys.} B}
\def\PLB{{\em Phys. Lett.}  B}
\def\PRL{\em Phys. Rev. Lett.}
\def\PRD{{\em Phys. Rev.} D}
\def\PR{{\em Phys. Rev.}}
\def\PRB{{\em Phys. Rev.} B}

\def\be{\begin{equation}}
\def\ee{\end{equation}}
\def\bea{\begin{eqnarray}}
\def\eea{\end{eqnarray}}

\begin{document}

\title{FERMION DAMPING RATE EFFECTS\\
 IN COLD DENSE MATTER \footnote{Talk given at ``Strong and 
Electroweak Matter 2000'', Marseille, June 2000.} 
\footnote{CERN-TH/2000-277}}

\author{Cristina Manuel}

\address{Theory Division, CERN, CH-1211 Geneva 23, Switzerland\\E-mail:Cristina.Manuel@cern.ch}


\maketitle\abstracts{We  review the non-Fermi or marginal liquid
behavior of a relativistic QED plasma. In this medium 
a quasiparticle has a damping rate that depends linearly
on the distance between its energy and the Fermi surface.
We stress that this dependence is due to the long-range character of the
magnetic interactions in the medium. Finally, we study how
the quark damping rate modifies the gap equation of color
superconductivity, reducing the value of the gap at the
Fermi surface.}

\section{Introduction}

There is an increasing interest in studying how matter behaves 
at very high density. While high density effects in non-relativistic
systems have been studied  thoroughly in the past, the  
same does not hold true for relativistic ultradegenerate plasmas.
Relativistic effects cannot be avoided if the chemical potential $\mu$
of the system is much larger than the mass of the
particles that form the medium. This situation certainly occurs
in the interior of neutron stars. The astrophysical scenario is
the natural domain of application of the physics of ultradegenerate
relativistic plasmas.

Electromagnetic plasmas  behave in a drastically different way 
in their non-relativistic and ultrarelativistic limits.
This is so because the magnetic interactions are suppressed in the
non-relativistic limit by powers of $v^2/c^2$, where $v$ is the 
typical velocity of the particles in the plasma, and $c$ is the
velocity of light. Electric and magnetic interactions behave in a 
very different way in a plasma. In the medium, static electric
fields are completely screened. This is the well-known Debye
screening phenomenon, also known as Thomas-Fermi screening for
ultradegenerate plasmas.  But magnetic interactions are only 
weakly dynamically screened, through Landau damping. Thus, while
electric interactions are short-ranged, magnetic interactions
are long-ranged. This fact has several relevant consequences for
ultradegenerate plasmas  and makes the relativistic
and non-relativistic phases of the plasma to look completely different.

In this talk  we will discuss how the long-range character of
magnetic interactions affects the lifetime of a quasiparticle in
the medium. This is based on work done in collaboration with
Michel Le Bellac \cite{LeBellac:1997kr,Manuel:2000mk}.
We will then see how the fermion lifetime effects also correct
the value of the gap of color superconductivity
\cite{Manuel:2000nh}.

\section{Lifetime of a quasiparticle and non-Fermi liquid 
behavior of the relativistic plasmas}

One of the central concepts in a plasma is that of a quasiparticle.
A particle immersed in a medium modifies its propagation
properties by interacting with the surrounding medium.
In  field theoretical language, we would say
that the particle is ``dressed" by a self-energy cloud. 
In the ultradegenerate plasma, the  relevant degrees of freedom 
are those of quasiparticles or quasiholes
(absences of particles in the Fermi sea) living close to the Fermi surface.
Because of the exclusion principle, quasiparticles/quasiholes can only live if 
they are outside/inside the Fermi sea. 
These excitations tend to lower their
energy by  undergoing collisions with the particles in the Fermi
sea. They  decay, and  thus have a finite lifetime.
The concept of quasiparticle,
however, only makes sense if its lifetime is long enough, or in other words, 
if its damping rate is much smaller than its energy.

If the interactions in the system are repulsive and short-ranged,
some of the propagation properties of the quasiparticles can be deduced
on general grounds. In that case, Luttinger's theorem \cite{Luttinger} states
the energy dependence of the damping rate of
a quasiparticle that lives close to the Fermi surface.
The damping rate can be obtained either by computing the imaginary
part of the fermion self-energy or, alternatively, by computing
the decay rate
\begin{eqnarray}
\label{decay}
\Gamma (E)& = & \frac{1}{E}  \int{{\rm d}^3p'\over (2\pi)^3}
\frac{ \left( 1- \Theta(\mu -E_{p'}) \right)}{2 E_{p'}}
\int{{\rm d}^3 k \over (2\pi)^3} 
\frac{  \Theta(\mu -E_{k})}{2 E_{k}}
\\ & \times & \int{{\rm d}^3 k' \over (2\pi)^3} 
\frac{ \left( 1- \Theta(\mu -E_{k'}) \right)}{2 E_{k'}} 
 (2 \pi)^4 \delta^{(4)} (P +K-P'-K') |{\cal M}|^2 \ ,
\nonumber
\end{eqnarray}
where $|{\cal M}|^2$ is the scattering matrix element squared,
and $\Theta$ is the step function.
The above decay rate represents the interaction of the quasiparticle
with one  fermion inside the Fermi sea with energy $E_k$.
As a result, two new particles appear, with energies $E_{k'}$ and $E_{p'}$,
which are outside the Fermi sea.    
If the interaction is repulsive and short-ranged, and for
$E-\mu \ll \mu$,
 one can take $|{\cal M}|^2$ with the value of 
the fermion energies at $\mu$. Then, 
one can deduce that
$\Gamma (E) \propto (E-\mu)^2$, only using the phase-space
restrictions of fermion-fermion scattering.

The damping rate can also be obtained from the
imaginary part of the fermion self-energy, and the computation
agrees with that obtained from Eq. (\ref{decay}).
Since the real and imaginary parts of the self-energy corrections
are related by dispersion relations, Luttinger's theorem implies that the
leading order behavior of the real part of the self-energy is
$ \propto |E-\mu|$. For weakly coupled systems, the dispersion
relations of the quasiparticles (or quasiholes) are not
drastically affected by medium effects.

For systems with long-range interactions, it is not possible
to make the previous general statements, as in general the integral in
Eq. (\ref{decay}) will depend  on the form of the interaction,
and in general, Luttinger's theorem will be violated. This is actually what 
happens in relativistic QED plasmas, due to the long-range
character of the magnetic interactions. A closer look into the
decay rate Eq. (\ref{decay}) in a relativistic plasma shows
that it is dominated by scattering in the forward or collinear
direction, mediated by a soft Landau damped magnetic photon.
The momentum of the photon in the process is space-like, so the
fermion damping rate would vanish in the absence of Landau
damping. In particular, one finds \cite{LeBellac:1997kr,Vanderheyden:1997bw} 
\begin{equation}
\label{imag}
{\rm Im} \Sigma_+ (E, p) \sim \frac{e^2}{24 \pi}  |E - \mu| \ ,
\end{equation}
where $e$ is the electromagnetic coupling constant. 
The real part of the self-energy can be computed from the
imaginary part, using a dispersion relation. One then finds
\cite{Brown:2000eh,Manuel:2000mk}
\begin{equation}
{\rm Re} \Sigma_+ (E, p) \sim \frac{e^2}{12 \pi^2} (E-\mu) \ln{
\frac{M}{|E-\mu|}} + {\cal O}((E-\mu)) \ . 
\end{equation}
The wavefunction renormalization factor $Z$ can then be equally
computed from the above values. One finds
\begin{equation}
Z^{-1} \sim 1 - \frac{e^2}{12 \pi^2} \ln{\frac{M}{|E-\mu|}} \ .
\end{equation}
Thus, in the limit $E \rightarrow \mu$, the fermion propagator
vanishes, instead of showing the typical step discontinuity 
associated to the existence of a Fermi surface \cite{Holstein}.
This is an anomalous behavior for a typical Fermi liquid. Its origin is
the long-range character of the magnetic interactions in the
relativistic system.

\section{Color superconductivity at weak coupling}

QCD at very high baryonic density behaves as a color superconductor
\cite{Rajagopal}.
This is a consequence of Cooper's theorem, as any attractive interaction
occurring close to the Fermi surface makes the system unstable to the
formation of particle pairing. In QCD the attractive interaction is
provided by one-gluon exchange in a color antisymmetric $\bar 3$ channel.

In the weak coupling limit the value of the gap  can be computed
in perturbation theory. The condensation process is dominated
by the exchange of very soft magnetic gluons, which are dynamically
screened by Landau damping. The Meissner effect is a subleading
effect in the gap equation. 
At leading order, one finds 
\cite{Son:1999,Schafer:1999b,Pisarski:2000bf,Pisarski:2000tv,Hong:2000tn,Hong:2000fh,Evans:1999at,Brown:2000aq,Brown:1999yd}
\begin{equation}
\label{eq.A13}
\phi_0 \sim 2 \, \frac{b_0}{g^5}\, \mu \exp{ \left(-\frac{3 \pi^2}{\sqrt{2}
  g}\right)}
\left[1 + {\cal O}(g) \right] \ , \qquad 
b_0 =  256 \pi^4 (\frac{2}{N_f})^{5/2} b'_0 \ , 
\end{equation}   
where $g$ is the gauge coupling constant,
 $N_f$ is the number of quark flavors, and $b'_0$ is a constant
of order one. The dependence on the coupling constant of the gap
is quite different from the one that arises
in a system with short-range interactions.

It is possible to compute  next to leading order corrections
to Eq. (\ref{eq.A13}) by introducing one-loop corrections in the quark
propagators of the gap equation. Here, we will mainly concentrate
on studying how the quark damping rate affects the value of the gap
close to the Fermi surface. At leading order one finds a modified
gap equation \cite{Manuel:2000nh}
\begin{equation}
\label{eq.A36}
\phi_{k} = \frac{ g^2}{36 \pi^2} \int^{\infty}_0  d (q-\mu)
\left[\ln{\left( \frac{\mu^2 b^2}{|\epsilon^2_q  -\epsilon^2_k|}\right)}
 \right] \frac{ \phi_{q}}{\epsilon_q} 
\frac{2}{\pi} \arctan{\left(\frac{\epsilon_q}{\Gamma_q}\right)} \ ,
\end{equation}
where $b=b_0/g^5$,
 $\epsilon_q=\sqrt{(q -\mu)^2 + \phi^2_q}$, and 
$\Gamma_q$ is the quark damping rate.
The most relevant effect of the damping
rate is introducing a physical ultraviolet cutoff in the gap equation:
when the ratio  $\epsilon_q/\Gamma_q$ starts to be small, the
integrand in Eq. (\ref{eq.A36}) tends to zero. This situation actually
occurs for quarks that are far away from the Fermi surface.
As expected, the condensation process only occurs close to the Fermi
surface.

For a leading order computation of the gap at the Fermi surface
one can take the value of
$\Gamma_q$ in the normal phase of the system (that is,
using Eq. (\ref{imag}), replacing $e^2$ by $\frac{4}{3}g^2$).
This is so because the one-loop fermion self-energy
in the normal phase differs from that in the
superconducting phase by, at most, 
values of the order of the squared of the condensate. Also,
the Meissner effect in the gluon propagator to arrive at the
value of $\Gamma_q$ is a subleading effect. The dominant scattering
processes are those occurring in the forward direction, and
these processes are dominated by soft Landau-damped color magnetic
interactions, exactly as it happens in the gap equation.
To leading order one finds \cite{Manuel:2000nh}
\begin{equation}
\label{eq.A42}
\phi_0^{damp} \sim 2 \, \frac{b_0}{g^5} \,\mu \exp{\left(- \frac{\pi}{2 
{\bar g}_{eff}} \right)} \ ,
\end{equation}
where ${\bar g}_{eff}^2 = {\bar g}^2 \left(1 - 2 {\bar g}^2 \right)$,
and $\bar g=g/3\sqrt{2}\pi$.
The value of the gap at the Fermi surface is then reduced.
This can be understood in very intuitive terms. The fact that the
quarks decay limits their efficiency to condense.
The decay of the quasiparticles should also affect the critical
temperature of transition to the normal phase of the system
computed in Refs. \cite{Pisarski:2000bf,Pisarski:2000tv,Brown:2000aq,Brown:1999yd}. 

The fermion damping rate effects represent a correction
of order $g^2$ to the leading order value Eq. (\ref{eq.A13}).
It is worth emphasizing that this is not the complete next-to-leading
order correction, which it  should be possible to compute using the 
Schwinger-Dyson equations.

\vskip 0.5cm

{\bf Acknowledgements}

I would like to thank H. Ren for very useful discussions on the
non-Fermi liquid behavior of the QED plasmas. My gratitude also
goes to the organizers of this nice meeting. Financial support
from a Marie Curie EC Grant (HPMF-CT-1999-00391) is 
acknowledged.

\end{document}